\documentclass[11pt,twoside]{article}
\usepackage{asp2004}
\usepackage{epsf}
\usepackage{psfig}
\begin{document}
\title{Beta, local SNIa data and the Great Attractor}
\author{John Lucey} 
\affil{Department of Physics, University of Durham, Durham, DH1 3LE, UK}
\author{David Radburn-Smith} 
\affil{Department of Physics, University of Durham, Durham, DH1 3LE, UK}
\author{Mike Hudson}
\affil{Department of Physics, University of Waterloo,
Waterloo, ON, CANADA, N2L 3G1}

\begin{abstract}
We compare the measured peculiar velocities of 98 local ($<150 h^{-1}$
Mpc) type Ia supernovae (SNIa) with predictions derived from the PSCz.
There is excellent agreement between the two datasets with a 
best fit $\beta_{\rm{I}}$ ($=\Omega_{\rm{m}}^{0.6}/b_{\rm{I}}$) of
$0.55\pm0.06$. Subsets of the SNIa dataset are further analysed
and the above result is found to be robust with respect to culls by
distance, host-galaxy extinction and to the reference frame in which
the analysis is carried out. 

We briefly review the peculiar motions in the direction of the
Great Attractor. Most clusters in this part of the sky 
out to a distance of 14,000 km\,s$^{-1}$, i.e. those closer 
than the Shapley Concentration, have sizable positive peculiar 
velocities, i.e. ($\sim$ +400 km\,s$^{-1}$). There are nine
local SNIa in the GA direction that are in the foreground of
Shapley. All these SNIa have positive peculiar velocities. 
Hence both the cluster and local SNIa data strongly support 
the idea of a sizable flow into Shapley.
\end{abstract}

\section{Introduction}
Peculiar motion studies are a powerful tool for examining the
underlying mass distribution of the local universe. 
In the linear regime, where density fluctuations are small, the mass
over-density $\delta_m$ can be related to the fluctuation in galaxy
number-density $\delta_g$ via $\delta_g(r)=b_{}\delta_m(r)$, where
$b$ is the linear bias parameter. This bias parameter together with
the cosmological mass density parameter $\Omega_m$ can be used to
predict peculiar velocity fields from all-sky redshift surveys.
The predicted velocities scale linearly with
the dimensionless quantity $\beta$ ($=\Omega_m^{0.6}/b$).
By comparing the peculiar velocities predicted by 
a galaxy density field with direct measurements, $\beta$ can be
determined.

All-sky galaxy samples derived from {\em{IRAS}} satellite data have been
used extensively to map the local density field. Currently the most
complete redshift survey of {\em{IRAS}} sources is provided by the
PSCz \citep{sau00}. This survey consists of redshifts for 15,411
galaxies uniformly distributed over 84.1\% of the sky with a median
redshift of 8,500 km s$^{-1}$. The PSCz survey's depth, excellent sky
coverage and density allow for the reliable mapping of the
distribution of galaxies in the local universe \citep{bra99}.
Typically velocity-velocity comparisons give values for 
$\beta_{\rm{I}}$ in the range 0.4 - 0.6 \citep[see][]{zar02}.

With distance errors less than $10\%$, Type Ia supernovae 
(SNIa) are an important probe of the local velocity field.
An early attempt to use SNIa was carried out by 
\citet{rie97} who compared the measured peculiar velocities of 24 SNIa 
to the velocities predicted 
from the 1.2~Jy {\em{IRAS}} redshift survey \citep{fis95} and the 
Optical Redshift Survey \citep{san95, bak98}. 
They derived $\beta_{\rm{I}}=0.4\pm0.15$ and
$\beta_{\rm{O}}=0.3\pm0.1$ respectively, with the relatively large error
resulting from the small sample size. 

Recently \citet{ton03} have produced a homogenized 
compendium of 230 SNIa for constraining cosmological quantities. 
The local ($<$\,150\,$h^{-1}$ Mpc) SNIa in this compendium are
a new valuable resource to investigate the local
peculiar velocity field. Here we discuss two applications:
the new measurement of $\beta_{\rm{I}}$ and the flow in the
direction of the Great Attractor.

\section{$\beta_{\rm{I}}$ determination}
 
We have compared the peculiar velocities measured from
this local SNIa sample to the peculiar
velocity field derived from the smoothed PSCz density field determined
by \citet{bra99}. In our analysis we only consider the 107 SNIa
that are closer than 150 $h^{-1}$ Mpc as the PSCz density field 
suffers from sizable shot noise at greater distances. 
The median distance error of 
this sample is $\sim8\%$.

\begin{figure}[!ht]
\hfil \psfig {figure=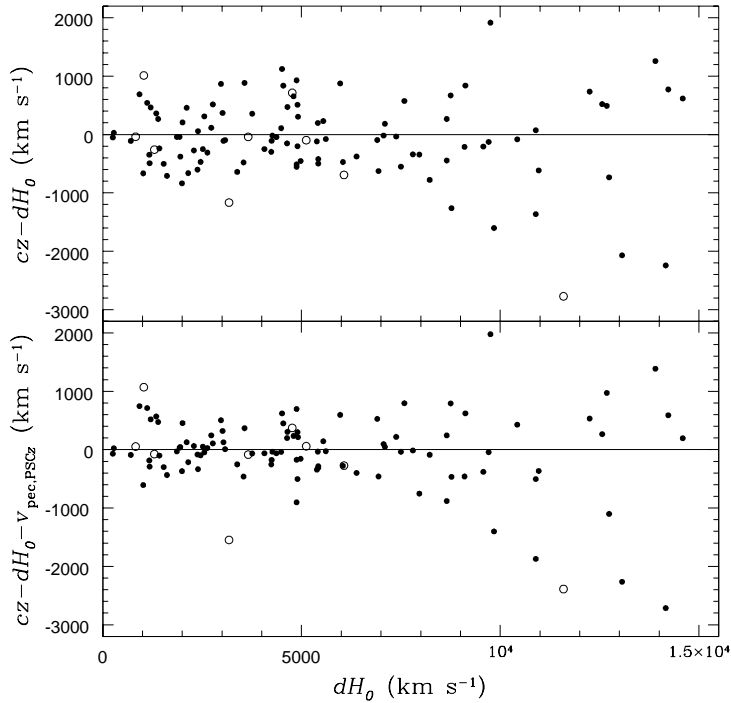,width=10cm} \hfil
\caption{
The Hubble flow for the 107 SNIa contained within 150 $h^{-1}$ Mpc in
the LG frame. 
The upper panel shows the original uncorrected data
whilst the lower shows the data after removal of the
$\beta_{\rm{I}}$\,=\,0.5 PSCz peculiar velocities. 
SNIa with host-galaxy extinctions
$A_V > 1.0$ are plotted as open circles.
}
\label{hubble}
\end{figure}

There is excellent agreement between the predicted PSCz 
peculiar velocities and those measured from the SNIa sample.
This is illustrated in Figure \ref{hubble} where the Hubble flow
for the local SNIa sample is shown. 
The lower panel of this figure has the
predicted PSCz velocities for the $\beta_{\rm{I}}$\,=\,$0.5$ case removed. 
In the range 20\,-\,80\,$h^{-1}$ Mpc, where the majority of 
SNIa lie, the rms scatter around the Hubble flow is reduced from 
490 to 390 km\,s$^{-1}$. In
Figure \ref{hubble}, nine SNIa with $A_V > 1.0$ are plotted
as open circles. Three of these nine SNIa are distinct outliers. Hence
in our analysis we have chosen to exclude the SNIa with $A_V>1.0$.

\begin{figure}[!ht]
\hfil \psfig {figure=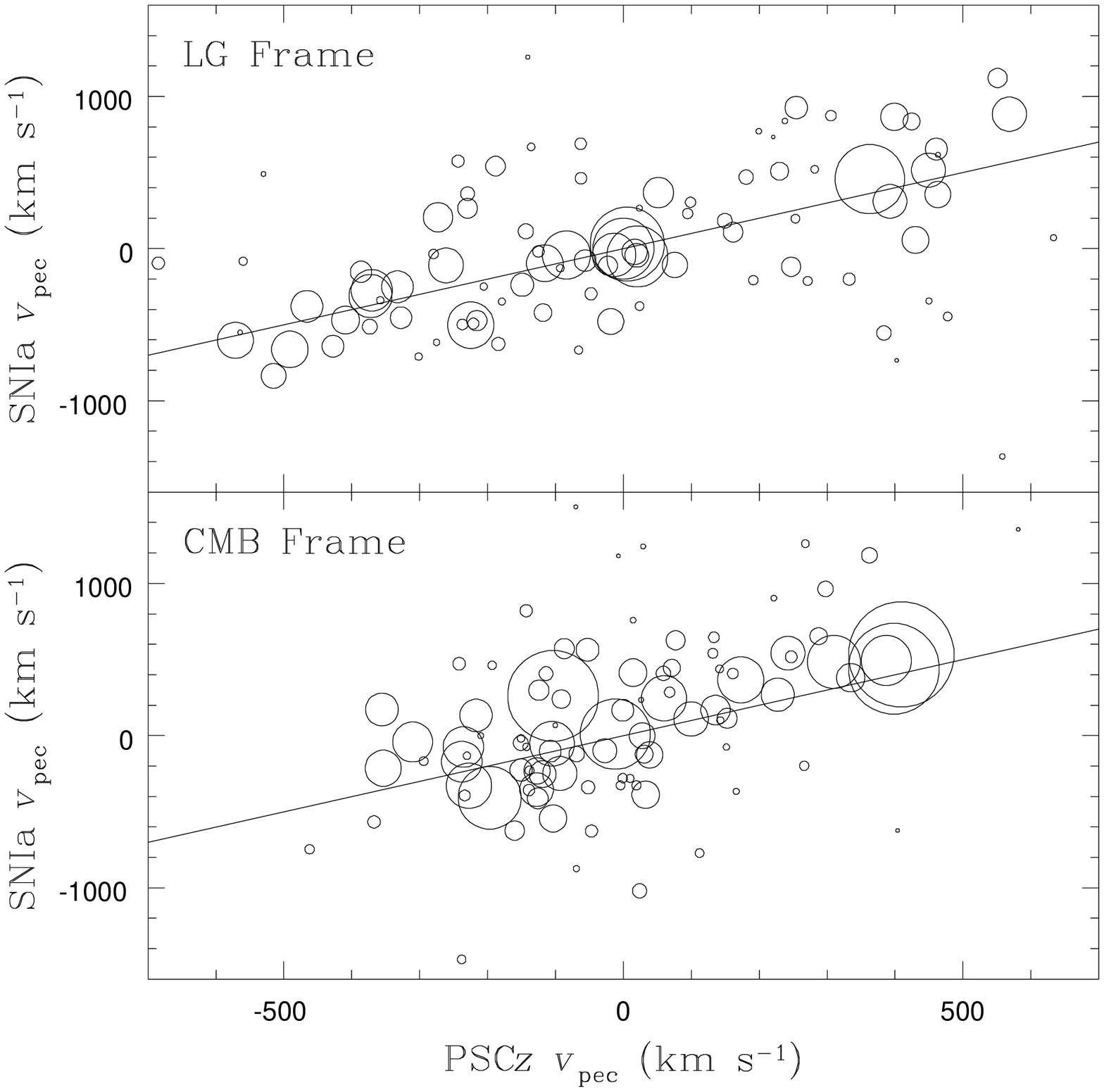,width=10cm} \hfil
\caption{
Comparison of SNIa peculiar velocities to PSCz predicted
    peculiar velocities in the range $0 h^{-1}$ Mpc to $150\,h^{-1}$ Mpc
    with $A_V < 1.0$ and $\beta_{\rm{I}}=0.55$. The size of the
    data point is inversely proportional to the total error
    ($\sigma=\sqrt{\sigma_{d}^2+\sigma_{cz}^2}$) on each SNIa. The
    smallest and largest circles correspond to values of
    $\sigma=$ 1290 km\,s$^{-1}$ and 170 km\,s$^{-1}$ respectively.
    The 1-to-1 line is shown in each panel. 
}
\label{fit}
\end{figure}

To determine $\beta_{\rm{I}}$ we minimise the $\chi^2$ relation:
\begin{equation}
  \chi^2=\sum_{i} \Biggl(\frac{(v_{i,{\rm{PSC_z}}}-v_{i,{\rm{SN}}})^2}{\sigma_{i,\rm d}^2+\sigma_{i,{\rm{cz}}}^2}\Biggr)
\label{eq:beta}
\end{equation}
where $v_i$ is the peculiar velocity of the $i^{th}$ supernova,
$\sigma_{\rm d}$ is the distance error and $\sigma_{\rm cz}$ incorporates
both an estimate of the error in redshift determination as well as an
additional dispersion factor. Our preferred value of $\sigma_{\rm cz}^{2}$
is $\sigma_{\rm{cl}}^2+150^2$ km\,s$^{-1}$, where $\sigma_{\rm{cl}}$
is an extra factor included for SNIa that lie near known rich clusters.
Undertaking the analysis in the Local Group (LG) frame with this choice of 
weighting, we derive a $\beta_{\rm{I}}$ of 0.55\,$\pm$\,0.06 with
a reduced $\chi^2_{\nu}$ of about 1 (see Figure \ref{fit} top panel). 

For other reasonable choices of  
$\sigma_{\rm cz}$ we find values for $\beta_{\rm{I}}$ in the range 0.54 to 0.57.
We have explored the robustness of these results
by considering various subsamples of the local SNIa dataset. 
We find that the derived $\beta_{\rm{I}}$ is insensitive 
to the distance range of the local SNIa considered, 
the $A_V$ cut adopted and the reference frame used for the
analysis, i.e. LG or CMB (Figure \ref{fit} lower panel). Hence the
measured peculiar velocities from the local SNIa sample are in very
good agreement with the peculiar velocities predicted from the PSCz
density field. A full account of this comparison is given in
\citet{rad04}.

\section{Great Attractor Flow}
The unexpected discovery by \citet{lyn88}
of a large ($\sim$600\,km\,s$^{-1}$)
outflow (positive peculiar velocities) in the Centaurus region
led to the concept of a large extended mass
distribution, nicknamed the Great Attractor (GA),
dominating the dynamics of the local universe.
Lynden-Bell et al. estimated that this structure was
located at ($l$, $b$, $cz$) $\sim$
(307$^\circ$, 7$^\circ$, 4,350\,$\pm$\,350\,km\,s$^{-1}$)
and had a mass of $\sim$5$\times$10$^{16}$\,M$_\odot$.
Despite over 15 years of study our understanding of
the GA is very limited. In particular the GA's extent and precise
location is still poorly known. For example, \citet{ton00}, using SBF
distances to 300 early-type galaxies, derived a much closer distance
for the GA, i.e. 
(289$^\circ$, 19$^\circ$, 3200\,$\pm$\,260\,km\,s$^{-1}$)
and a mass of $\sim$8$\times$ 10$^{15}$\,M$_\odot$, i.e. a factor of
$\sim$6 less than the original GA value. Alternatively \citet{wou99}
have argued that the very rich Norma cluster  (Abell 3627) at
(325$^\circ$, --7$^\circ$, 4,800\,km\,s$^{-1}$) may mark the ``core''
of the GA.

Attempts to measure the expected GA backside infall have proved
controversial \citep[see][]{mat92}. Some studies have argued for a
continuing high amplitude flow beyond the GA distance resulting from
the gravitational pull of the Shapley Concentration
\citep{sca89,all90,hud99,bra99}. The Shapley Concentration is a
remarkably rich concentration of galaxies in northern Centaurus
\citep[see][]{rei02}. This structure contains more than 20 Abell/ACO
clusters within 25\,h$^{-1}$ Mpc of the very rich A3558 cluster at
(312$^\circ$, 31$^\circ$, 14,500\,km\,s$^{-1}$). The GA, in contrast,
only contains at most five rich clusters.

Shapley's contribution to the Local Group's motion is unclear as it
is difficult to decouple this from the flow towards the GA. 
\citet{bar00} estimated that Shapley was responsible for 
only 26\,km\,s$^{-1}$ of the Local Group's motion with 
respect to the CMB. For the HST key project, \citet{mou00} adopted a 
value for this component of 85\,km\,s$^{-1}$. 
From peculiar velocities of 10 clusters in this region,
\citet{smi99} found that both the GA and Shapley generate 
50\,$\pm$\,10\% of the Local Group motion. Analysis of the local tidal
field implied a value in the range 100 to 200\,km\,s$^{-1}$
\citep{hof01}. Recent observations of the X-ray dipole \citep{koc04}
have implied an even larger contribution from Shapley.

\begin{figure}[!ht]
\hfil \psfig {figure=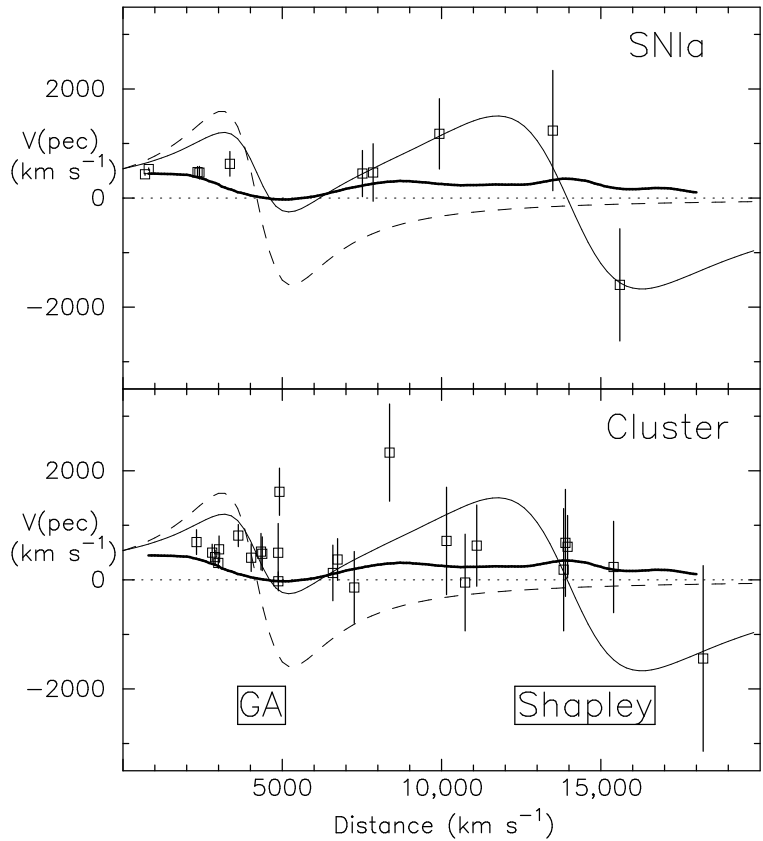,width=10cm} \hfil
\caption{
The cluster peculiar velocity data in the GA region from 
SMAC, ENEARc, SCI, SCII and SBF surveys. 
The central bold curve shows the PSCz predictions in this direction. 
Most data points lie above this curve
suggesting that the resolution of the sparse PSCz sampling has been unable
to reliably measure the high amplitude of the flow in this direction.
The other two curves are from 
Faber-Burstein models which are normalised to produce the 
Local Group's CMB motion; the one attractor model is centred on 
the GA at $\sim$4350 km\,s$^{-1}$ (dashed curve); the two 
attractor model has the GA
and Shapley Concentration equally contributing to 
the Local Group's CMB motion (thin line curve).
In each case the maximum amplitude of the model is shown.
}
\label{ga_flow}
\end{figure}

There are now several peculiar velocity surveys that include
measurements in the GA region, i.e.
the Tully-Fisher work of SCI/SCII \citep{gio99},
the Fundamental Plane work of SMAC {\citep{hud04},
the $D_n-\sigma$ work of ENEARc \citep{ber02}
and the SBF survey \citep{ton00}.
Following \citet{smi99}, in Figure \ref{ga_flow} 
(lower panel) we show the cluster peculiar 
velocities for the GA/Shapley direction from the above sources
together with model predictions. 
While the large random errors limit our resolution of the 
flow pattern, there are a number 
of features that are apparent: 
(a) most GA clusters (cz $<$ 5,000 km\,s$^{-1}$), i.e.
those traditionally associated with the GA, 
have positive peculiar velocities of $\sim$400 km\,s$^{-1}$;
(b) clusters that lie immediately beyond the GA have small
peculiar velocities and do not display any evidence of backside infall
towards the GA; (c) clusters in the range 80 to 120\,$h^{-1}$ Mpc have
positive peculiar velocities but the significance of this result is low.
The original GA model is clearly a very poor fit to the
cluster data. A two attractor model (GA and Shapley Concentration) 
successfully accounts for the lack of observed GA backside infall.

In the GA region there are 10 SNIa and these are plotted in
Figure \ref{ga_flow} (upper panel). The nine SNIa closer than
Shapley all have positive peculiar velocities.
Hence both the cluster and SNIa observations provide
strong support for the presence of a sizable flow towards Shapley.
Unfortunately the random errors on the peculiar velocities 
for objects in the crucial zone between the GA and Shapley are 
large and this severely limits how well the GA/Shapley mass
ratio can be determined.
New surveys, i.e. NOAO Fundamental Plane Survey \citep{smi04}, 
and future observing programmes have the potential to resolve
this important issue.
 
\acknowledgements 
The authors would like to thank Enzo Branchini for providing the PSCz
velocity field. DJR-S thanks PPARC for a
research studentship. MJH acknowledges support from NSERC and the ORDCF.

\end{document}